\documentclass[aps,prb,showpacs]{revtex4}

\usepackage{amsfonts,amsmath,amssymb}
\usepackage{graphicx,color}
\usepackage{verbatim}

\begin{document}

\title{Simulation of the thermally induced austenitic phase transition in NiTi nanoparticles}

\author{Daniel Mutter}
\email{daniel.mutter@uni-konstanz.de}
\affiliation{Department of Physics, University of Konstanz, 78457 Konstanz, Germany}
\author{Peter Nielaba}
\affiliation{Department of Physics, University of Konstanz, 78457 Konstanz, Germany}

\date{\today}

\begin{abstract}
The reverse martensitic (``austenitic'') transformation upon heating of equiatomic nickel-titanium nanoparticles with diameters between 4 and 17 nm is analyzed by means of molecular-dynamics simulations with a semi-empirical model potential. After constructing an appropriate order parameter to distinguish locally between the monoclinic B19$'$ at low and the cubic B2 structure at high temperatures, the process of the phase transition is visualized. This shows a heterogeneous nucleation of austenite at the surface of the particles, which propagates to the interior by plane sliding, explaining a difference in austenite start and end temperatures. Their absolute values and dependence on particle diameter are obtained and related to calculations of the surface induced size dependence of the difference in free energy between austenite and martensite.
\end{abstract}


\maketitle

\section{Introduction}
\label{sec1}
Shape-memory alloys (SMAs) exhibit the unique effect of ``remembering'' their original shape after deformation and heating [``shape memory effect'' (SME)], and moreover, they show ``super-'' or ``pseudoelastic'' behavior (SE), i.e. recoverable strains of up to 8\%.\cite{buchots:98} These features make them to widely used and studied materials in industry and research. The underlying physical effect is the martensitic phase transition (MT), where a crystal lattice transforms diffusionless between a high-symmetric structure (``austenite'') and a low-symmetric structure (``martensite''), depending on external temperature or stress. Within the scope of nanoscience and -technology, extensive effort has been made in the last decade to explore MT, SME, and SE at the nanoscale (see Waitz \emph{et al.} \cite{waitsu:09} for an actual overview). Especially in the case of near-equiatomic nickel-titanium, the most frequently used SMA, many experimental studies examine the behavior of different nanostructures such as particles \cite{waispi:05,waipra:08,waiant:08,liucan:06,cascue:08,fushe:04}, thin films \cite{wankom:05,fuzha:06} or pillars \cite{yemis:10}. In most of these works, the dependence of system size on the structural transformation temperatures or stresses on the one hand, and on resulting morphologies on the other hand is discussed, as well as the question, whether there is a critical diameter, below which no MT can occur.\\
\indent In order to analyze this kind of questions from a theoretical point of view, molecular dynamics (MD) simulations have been performed in the past for different nanostructured alloys and pure metals, which can also show MT. Examples are simulations of Cu films \cite{kolgun:08}, NiAl wires \cite{par:06}, and CdSe particles \cite{grudel:09}, where stress induced transitions take place, or of FeNi particles \cite{kadgru:03} and Fe wires \cite{sanurb:09.2} in the case of thermally induced MT. Visualization of the atomistic transformation behavior of these systems has shown, that mainly free surfaces act as nucleation sites for the emerging structures, suggesting an explanation of decreasing transition temperatures with system size by an increasing surface-to-volume fraction.\\
\indent This work focuses on the thermally induced reverse martensitic (``austenitic'') phase transition in NiTi nanoparticles, occurring between the monoclinic B19$'$ at low temperatures and the cubic B2 structure at high temperatures. MD is applied with a Finnis-Sinclair potential \cite{finsin:84}, with NiTi parameters determined by Lai and Liu \cite{lailiu:00}. In a recent work \cite{mutnie:10}, the authors of the present paper applied these methods successfully to bulk NiTi, where a strong dependence of transition temperatures on Ni concentration has been found and could be explained by energetic calculations. The present work extends these studies to equiatomic NiTi nanoparticles, by visualizing the transformation process, and by determining for the first time absolute values of austenitic transition temperatures of NiTi as a function of system size. As in the case of concentration variation, this dependence can be explained by the change of the energy of the system, in this case due to free surfaces. The paper is organized as follows: In Sec.\:\ref{sec2}, a detailed structural analysis is presented and an order parameter is defined, which allows to distinguish between the occurring structures. In addition, details of the simulation method are given. The phase transition upon heating of the particles is analyzed and discussed in Sec.\:\ref{sec3}. Finally, a conclusion is given in Sec.\:\ref{sec4}.

\section{Structural analysis}
\label{sec2}
\begin{figure}
\resizebox{0.6\columnwidth}{!}{
\includegraphics{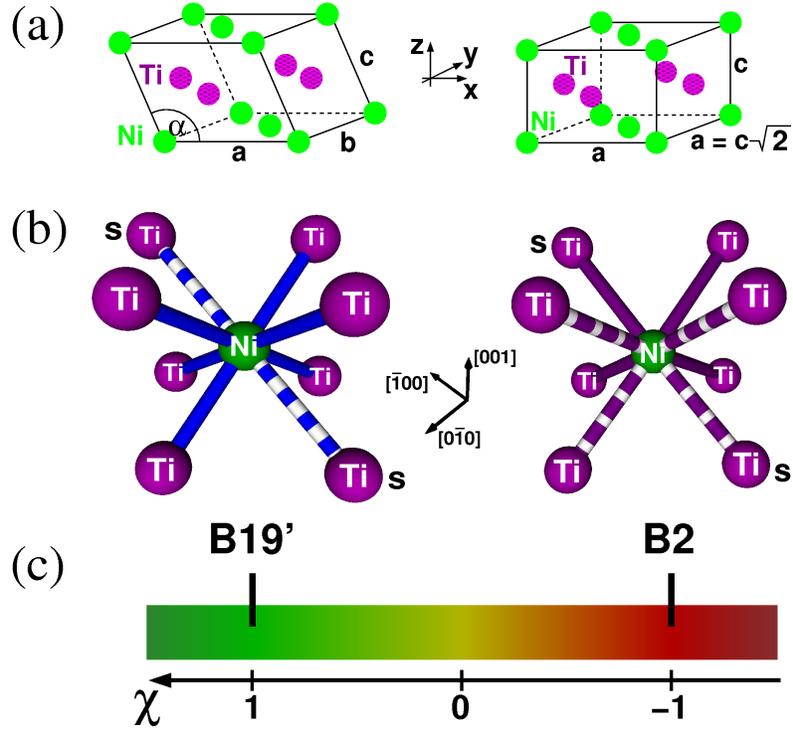}
}
\caption{To distinguish locally between B19$'$ (left half) and B2 (right half): (a) schematic representation of the lattice cells; (b) nearest neighbor environments with atoms in shear direction denoted by s and neighbor distances $2.85\ \mathring{\mbox{A}}$ (B19$'$, dashed), $2.51\ \mathring{\mbox{A}}$ (B19$'$, continuous), $2.69\ \mathring{\mbox{A}}$ (B2, dashed), $2.54\ \mathring{\mbox{A}}$ (B2, continuous); (c) resulting value of the order parameter $\chi$ and a translation to a color code}
\label{f1}
\end{figure}

As mentioned above, it has been shown recently \cite{mutnie:10}, that a slight modification of the cut-off behavior of effective hopping integrals and pairwise interactions leads to a stable B19$'$ structure with improved lattice parameters $a$ = $4.45\ \mathring{\mbox{A}}$, $b$\,=\,$4.03\ \mathring{\mbox{A}}$, $c$\,=\,$3.00\ \mathring{\mbox{A}}$, and $\alpha$\,=\,$97.7^{\circ}$ of the monoclinic cell at $T$\,=\,$0$ K in bulk systems, if compared to experimental \cite{prokor:04} and \emph{ab initio} \cite{hatkon:09.2} values (see Fig.\:\ref{f1}(a) for a schematic structure representation). The body-centered orthorhombic structure B33, proposed theoretically by Huang et al. \cite{huaack:03} to be the lowest energy state of NiTi with a shear angle of 107$^{\circ}$ is not reproduced by the choice of potential parameters used in this work. Until now, B33 could not be stabilized experimentally in pure NiTi, neither in bulk nor in nanoscaled systems, which is assumed to be due to internal stresses, leading to the monoclinic B19$'$ as the martensitic structure \cite{huaack:03}. Besides B19$'$ at low temperatures, the potential used in this work predicts a phase transformation to a cubic structure close to B2 ($c$\,=\,$3.01\ \mathring{\mbox{A}}$) if the system is heated above the austenitic transition temperature $T_A$\,=\,318 K in aMD simulation under periodic boundary conditions \cite{mutnie:10}. Since the reverse transition upon cooling produces a related, but different crystal structure, which is energetically slightly more favorable than B19$'$, the present work focuses only on the austenitic phase transition process. The nanoparticles are realized as spheres under free boundary conditions. MD is performed by a velocity-Verlet algorithm \cite{swoand:82} (timestep $\Delta t$\,=\,$10^{-15}$ s) with temperature control by a Nos\'{e}-Hoover thermostat \cite{nos:84}.\\
\indent Since it is not possible to identify the occurrence of the structural phase transition by changes of the shape of a simulation box when applying free boundary conditions, an order parameter is necessary to distinguish locally between B19$'$ and B2. Therefore, the nearest neighbor environments of these two structures are considered, as they are produced by the potential in a bulk simulation. In B19$'$ [Fig.\:\ref{f1}(b), left], the dashed lines symbolize the neighbor distances, which are elongated about 13.5\% relative to the continuous ones due to the monoclinic shear in $\left[\bar{1}00\right]$ direction. In B2 [Fig.\:\ref{f1}(b), right], the formerly two sheared lengths have nearly the same values as the unsheared ones [there is a difference in neighbor lengths of $\pm$\,3\% relative to a perfect B2 NiTi ($2.62\ \mathring{\mbox{A}}$), which is denoted by the dashed and continuous lines in this case]. By denoting the mean value of the two lengths in $\left[\bar{1}00\right]$ direction as $d_1$ and the mean value of the six lengths in the other directions as $d_0$, one can construct a quantity $\chi$\,=\,$\chi\left(d_0,d_1\right)$, which shall be equal to 1, if $d_1$\,=\,$d_1^{B19'}$\,=\,$2.85\ \mathring{\mbox{A}}$ and $d_0$\,=\,$d_0^{B19'}$\,=\,$2.51\ \mathring{\mbox{A}}$, and equal to -1, if $d_1$\,=\,$d_1^{B2}$\,=\,$d_0$\,=\,$d_0^{B2}$\,=\,$2.62\ \mathring{\mbox{A}}$. A linear ansatz $\chi\left(d_0,d_1\right)$\,=\,$A_0d_0 + A_1d_1$ leads to:
{\small
\begin{equation}
\chi\left(d_0,d_1\right) = \frac{d_0\left(d^{B2}+d_1^{B19'}\right)-d_1\left(d^{B2}+d_0^{B19'}\right)}{d^{B2}\left(d_0^{B19'}-d_1^{B19'}\right)}.
\label{g0}
\end{equation}
}
This order parameter is capable of distinguishing between B19$'$ and B2 on an atomic scale and of resolving intermediate states, but it is restricted to single variant domains with known shear direction, as studied in this work. By looking at systems with differently orientated monoclinic variants, one has to search for the two opposite elongated next neighbor distances in B19$'$ for each atom first, before applying Eq.\:\ref{g0}.
\\
\indent In order to simulate the austenitic phase transition in equiatomic NiTi nanoparticles, they are set up as spheres with diameters $d$ ranging from 4 to 17 nanometers in a perfect B19$'$ single variant structure, corresponding to a number of atoms between about 2000 and 196000. In this size range, the single B19$'$ structure in the nanoparticles is stable, but since the total energy of the surface atoms in B2 structure is lower than in B19$'$ orientation, this is only valid below a critical surface-to-volume ratio corresponding to a minimal size. Below this size, the particle contains of B2 even at low temperatures. After an equilibration time of 40000 timesteps, where the imposed temperature is reduced stepwise to 1 K, the systems are heated about 2 K every 2000 simulation steps up to 350 K. Heating rates in the range of 1 K/ps are usual in MD simulations of solid-solid phase transitions \cite{kadent:02,meyent:98}, and can even be realized in experiments by laser irradiation and shock-wave loading \cite{luoahr:03,luoahr:03.2}. By performing MD, Sandoval and Urbassek showed, that there is only a weak dependence of transition temperatures of Fe nano wires on the heating/cooling rate \cite{sanurb:09.2}.

\section{Results and discussion}
\label{sec3}
\begin{figure}
\resizebox{0.6\columnwidth}{!}{
\includegraphics{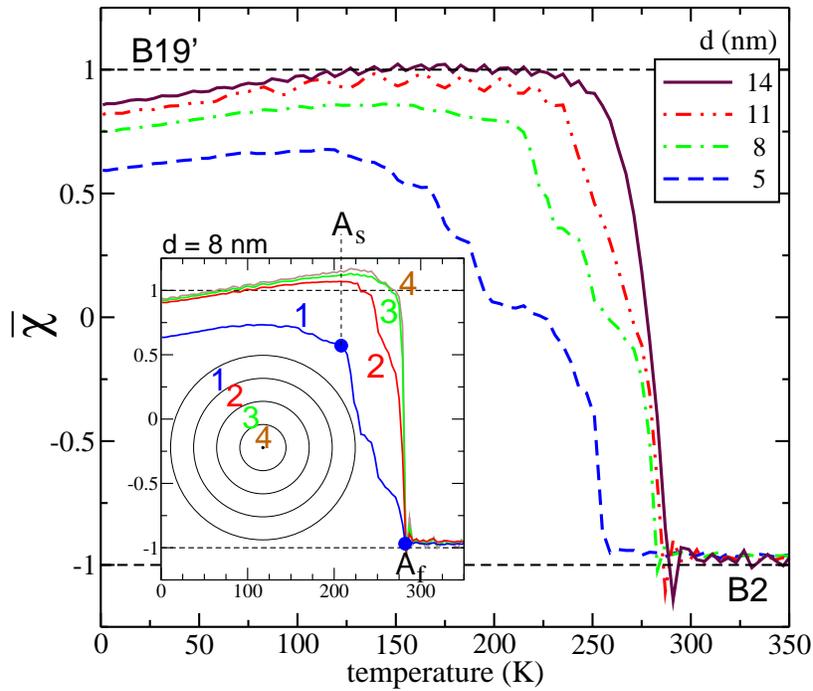}
}
\caption{Mean value of the structural order parameter $\chi$ of all atoms during a heating process of nanoparticles with different diameters $d$. Inset: Mean values of $\chi$ in spherical shells with a thickness of 1 nm.}
\label{f2}
\end{figure}

\begin{figure}
\resizebox{0.6\columnwidth}{!}{
\includegraphics{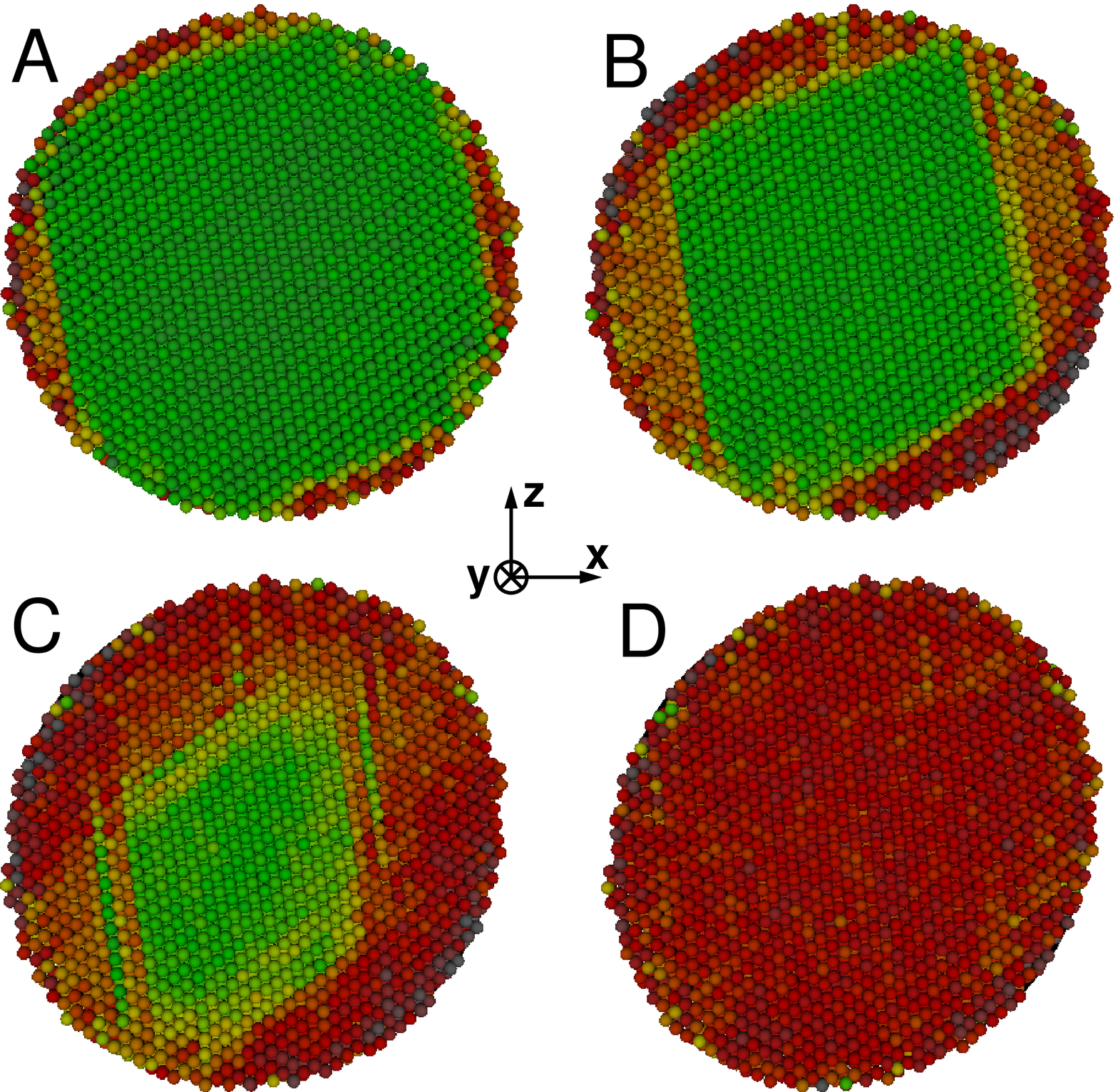}
}
\caption{Visualization of a transforming nanoparticle with diameter $d$\,=\,10 nm. The xz-plane through the origin is shown at temperatures (A) 239 K, (B) 259 K, (C) 275 K, and (D) 289 K.}
\label{f3}
\end{figure}
 Fig.\:\ref{f2} shows the mean $\chi$-value of all atoms during the heating process at 4 different system sizes. The smaller the particle, the more atoms lie at the surface, leading to the deviation of $\overline{\chi}$ in the martensitic state from the bulk value $\overline{\chi}$\,=\,1, since the surface atoms do not ``see'' a perfect  B19$'$ structure around them. At first, $\overline{\chi}$ increases and can obtain even values above 1. This is due to thermal expansion of the next neighbor distances leading to the atomic $\chi$-values (see Eq.\:\ref{g0}), out of which $\overline{\chi}$ is calculated. Nevertheless, the structures remain stable until a certain temperature is reached, where a transformation to the B2 phase ($\overline{\chi}$\,=\,-1) starts. The difference in austenitic start and end temperatures (called $A_s$ and $A_f$) indicates a heterogeneous transformation, starting at certain nucleation sites and propagating through the system. Since this difference depends on particle size, a surface triggered transformation can be assumed. Therefore it is necessary to use a more local order parameter instead of the global $\overline{\chi}$, where the core atoms are not taken into account, to identify the onset of the structural change more precisely. To this end, the nanoparticles are subdivided into spherical shells with a thickness of 1 nm, and the mean $\chi$ in each of these shells is calculated. For the particle with $d$\,=\,8 nm, the result is shown in the inset of Fig.\:\ref{f2}, where the different temperatures, at which the austenite reaches the inner lying shells, are visible. With this method, $A_s$ can be determined quantitatively for all system sizes, as the temperature, at which a significant drop of the structural order parameter of the outermost shell occurs. $A_f$ is defined as the temperature, where the $\overline{\chi}$-value of the innermost shell reaches -1. Values $\overline{\chi}$\,$<$\,-1 in the vicinity of $A_f$ are due to non-equilibrium states slightly deviating from B2.\\
\indent In order to understand the atomistic details of the transformation behavior, a visualization of the system is done by relating the atomic $\chi$ values to colors [Fig.\:\ref{f1}(c)]. Fig.\:\ref{f3} shows a central plane of a nanoparticle with diameter $d$\,=\,10 nm at four different temperatures between $A_s$ and $A_f$. The austenite transformation starts at the surface and propagates through the particle by sliding of certain lattice planes. If the particle is visualized not only in the xz-plane, but also in the xy- and yz-planes in the same manner, a martensite polyhedron can be identified, which is surrounded by $\left(100\right)$, $\left(11\bar{1}\right)$, $\left(\bar{1}11\right)$, $\left(111\right)$, and $\left(1\bar{1}1\right)$ lattice planes. This behavior can be understood as follows. The atoms at the surface of a spherical nanoparticle are arranged in many differently orientated planes. Each of these free surface planes has a certain energy difference per atom between martensite and austenite ($\Delta e$). If the particle is heated, the atoms at planes with the smallest $\Delta e$ values will change their structure first, at $T$\,=\,$A_s$. Table\:\ref{tab1} shows these differences for certain lattice planes, calculated separately by setting up systems in B19$'$ and B2 structure with free surfaces of the desired orientation at $T$\,=\,0 K, equilibrating them in a simulation, and averaging the energies of the surface atoms.
\begin{table}
\caption{Differences in surface energy per atom between B19$'$ and B2 [$\Delta e^{\left(hkl\right)}$\,=\,$e^{\left(hkl\right)}(\mbox{B19}')-e^{\left(hkl\right)}(\mbox{B2})$] for some lattice planes $\left(hkl\right)$ (in eV).}
\label{tab1}
\begin{tabular}{l l}
\hline\noalign{\smallskip}
Lattice planes $\left(hkl\right)$&$\Delta e^{\left(hkl\right)}$\\
\noalign{\smallskip}\hline\noalign{\smallskip}
$\left(010\right)$,$\left(111\right)$,$\left(1\bar{1}1\right)$&-0.06\\
$\left(100\right)$,$\left(\bar{1}11\right)$,$\left(11\bar{1}\right)$&-0.12\\
$\left(001\right)$,$\left(011\right)$,$\left(0\bar{1}1\right),\left(110\right)$,$\left(101\right)$,$\left(\bar{1}10\right),\left(\bar{1}01\right)$&-0.17\ \ -\ \ -0.21\\
\noalign{\smallskip}\hline
\end{tabular}
\end{table}
\\
In general, energetic barriers have to be overcome in martensitic transforming materials to enable the first order phase transitions. These barriers result from nonchemical energy contributions as elastic strain energies or unlike interfaces, which necessarily have to emerge, if a certain crystal structure forms out of another one \cite{buchots:98}. This barrier, which prevents a transformation below a certain temperature in the bulk like interior of the particle, is lowered for the atoms lying at the interfaces between the already transformed surfaces and the bulk. This leads to a structural change of these atoms at a lower temperature compared to the bulk value of $A_s$, what is proceeded plane by plane, until a temperature slightly below $A_f$ is reached. Here, the remaining martensite transforms homogeneously to B2, which can also be seen in the $\bar{\chi}$-curves of Fig.\:\ref{f2}, where the slopes drop nearly vertically before the transformation is completed. Going to larger systems, the influence of the surface shrinks, which implies a more homogeneous transformation and a less pronounced lattice plane sliding, resulting in a decreasing difference between $A_s$ and $A_f$.

\begin{figure}
\resizebox{0.6\columnwidth}{!}{
\includegraphics{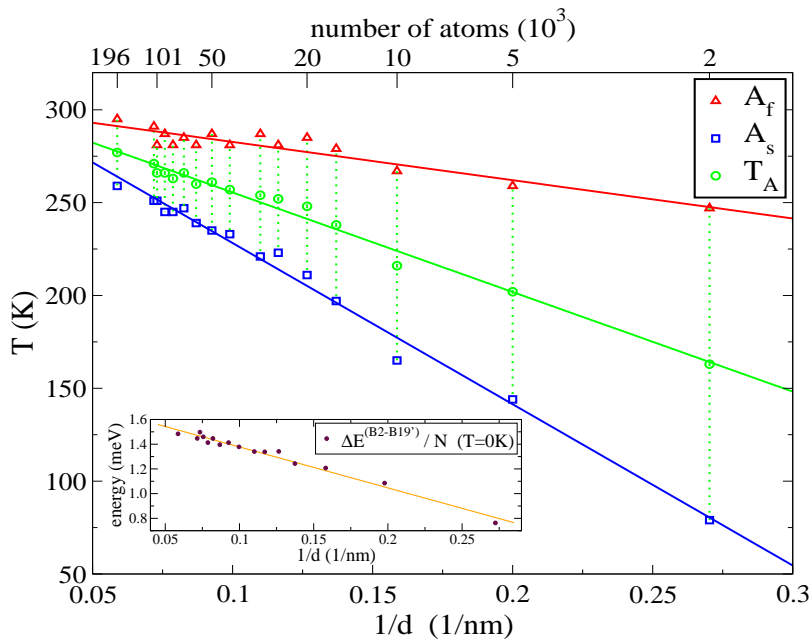}
}
\caption{Dependence of austenitic transformation temperatures on particle diameter and the total number of atoms. Inset: difference between austenite and martensite energy per atom as a function of 1/d. Lines represent linear data fits.}
\label{f4}
\end{figure}

\indent Plotting the simulation results of $A_s$ and $A_f$ as well as the mean austenitic transition temperature $T_A$\,=\,$0.5$ $\left(A_s+A_f\right)$ against $1/d$ (see Fig.\:\ref{f4}) leads to the behavior
\begin{equation}
T_A = T_A^{\mbox{\scriptsize bulk}} - \frac{\beta}{d}
\label{g2}
\end{equation}
by linear regression, with parameters $\beta$\,=\,536.3 nm$\cdot$K, and $T_A^{\mbox{\scriptsize bulk}}$\,=\,309.11 K. Thermodynamic calculations of different nano particles, which undergo structural phase transitions, have shown, that in these systems, the difference in free energy per atom between the two involved structures ($\Delta f^{a-m}$) shows a proportionality with $1/d$, too \cite{qinche:01,menzho:02}. Since this result is obtained by separating $\Delta f^{a-m}$ into a bulk and a surface part, it can be assumed, that it is the surface induced size dependence of $\Delta f^{a-m}$, which determines the $1/d$ behavior of the transition temperatures. For the NiTi nanoparticles, a calculation of $\Delta f^{a-m}$ at $T$\,=\,0 K, where $\Delta f^{a-m}$ equals the difference in internal energy per atom, has been carried out, which confirms the linear dependence on $1/d$ (see inset in Fig.\:\ref{f4}). The results are obtained by setting up different sized spheres in B19$'$ and B2 structure, and calculating the mean energy per atom after equilibration at T\,=\,0 K.
\\
\indent The $T_A^{\mbox{\scriptsize bulk}}$ value of Eq.\:\ref{g2} fits well with $T_A$\,=\,318 K, obtained by periodic boundary conditions \cite{mutnie:10}. Because in that case, the transformation is completely homogeneous, no difference between $A_s$ and $A_f$ could be identified. But since real NiTi systems always contain free surfaces in the form of grain boundaries, as well as other lattice defects or dislocations, which can act as nucleation sites, one should expect a non-zero $\Delta_A$\,=\,$A_f-A_s$, what in fact is observed experimentally in bulk NiTi with Ni concentration between 50.3\% and 51\%, lying in the range of 29-45 K \cite{khaami:09}. This is in good agreement with the values obtained for nanoparticles bigger than $d$\,$\approx$\,$11$ nm (50000 atoms), for which $\Delta_A$ seems to stay within about $37\pm5$ K. Since the absolute experimental value of $T_A$ (331 K for Ni$_{50.3}$Ti$_{49.7}$) \cite{khaami:09} fits well with $T_A^{\mbox{\scriptsize bulk}}$, it is assumed, that overheating effects due to the strong heating rate of 1 K/ps are negligible.\\
\indent Eq.\:\ref{g2} was observed in other nanosized systems and materials showing structural phase transitions, too, such as Co particles \cite{wenhua:06}, FeNi particles \cite{kadgru:03} or Fe nanowires \cite{sanurb:09.2}. In general, the proportionality of transformation temperature to the inverse diameter in small systems, which undergo first order phase transitions is known as the ``thermodynamic size effect''. It was derived at first by Pawlow in 1909 for the solid-liquid transition in small particles by a pure thermodynamic calculation of the gas pressure at a given particle size and temperature \cite{paw:09}.

\section{Conclusion}
\label{sec4}
In conclusion, an analysis of the austenitic phase transition in spherical NiTi nanoparticles is presented. By calculating next neighbor lengths, an order parameter is constructed, which can resolve changes between the low temperature structure B19$'$ and the high temperature structure B2 at an atomic scale. It is shown, that there is a heterogeneous structural transformation from the outer to the inner lying atoms of the particles upon heating above the austenitic start temperature $A_s$. Visualization of the transformation process shows a sliding of martensite/austenite interfaces of certain orientation, triggered by unequal energy differences between B19$'$ and B2 configurations at different free surface planes. Transition temperatures as a function of system size are obtained, leading to a $T_A$\,$\propto$\,$d^{-1}$ law, which is well established in nano science. This behavior is related to the size dependent energetic difference between austenite and martensite, triggered by the surface-to-volume ratio of the nanoparticles.

\begin{acknowledgments}
We gratefully acknowledge the support of the SFB 767, the NIC, and the HLRS.
\end{acknowledgments}


\begin{thebibliography}{32}
\expandafter\ifx\csname natexlab\endcsname\relax\def\natexlab#1{#1}\fi
\expandafter\ifx\csname bibnamefont\endcsname\relax
  \def\bibnamefont#1{#1}\fi
\expandafter\ifx\csname bibfnamefont\endcsname\relax
  \def\bibfnamefont#1{#1}\fi
\expandafter\ifx\csname citenamefont\endcsname\relax
  \def\citenamefont#1{#1}\fi
\expandafter\ifx\csname url\endcsname\relax
  \def\url#1{\texttt{#1}}\fi
\expandafter\ifx\csname urlprefix\endcsname\relax\def\urlprefix{URL }\fi
\providecommand{\bibinfo}[2]{#2}
\providecommand{\eprint}[2][]{\url{#2}}

\bibitem[{\citenamefont{1}}]{buchots:98}
\bibinfo{editor}{\bibfnamefont{K.}~\bibnamefont{Otsuka}} \bibnamefont{and}
  \bibinfo{editor}{\bibfnamefont{C.~M.} \bibnamefont{Wayman}}, eds.,
  \emph{\bibinfo{title}{Shape Memory Materials}} (\bibinfo{publisher}{Cambridge
  University Press}, \bibinfo{year}{1998}).

\bibitem[{\citenamefont{2}}]{waitsu:09}
\bibinfo{author}{\bibfnamefont{T.}~\bibnamefont{Waitz}},
  \bibinfo{author}{\bibfnamefont{K.}~\bibnamefont{Tsuchiya}},
  \bibinfo{author}{\bibfnamefont{T.}~\bibnamefont{Antretter}},
  \bibnamefont{and} \bibinfo{author}{\bibfnamefont{F.~D.}
  \bibnamefont{Fischer}}, \bibinfo{journal}{MRS BULLETIN}
  \textbf{\bibinfo{volume}{34}}, \bibinfo{pages}{814} (\bibinfo{year}{2009}).

\bibitem[{\citenamefont{3}}]{waispi:05}
\bibinfo{author}{\bibfnamefont{T.}~\bibnamefont{Waitz}},
  \bibinfo{author}{\bibfnamefont{D.}~\bibnamefont{Spisak}},
  \bibinfo{author}{\bibfnamefont{J.}~\bibnamefont{Hafner}}, \bibnamefont{and}
  \bibinfo{author}{\bibfnamefont{H.~P.} \bibnamefont{Karnthaler}},
  \bibinfo{journal}{Europhys. Lett.} \textbf{\bibinfo{volume}{71}},
  \bibinfo{pages}{98} (\bibinfo{year}{2005}).

\bibitem[{\citenamefont{4}}]{waipra:08}
\bibinfo{author}{\bibfnamefont{T.}~\bibnamefont{Waitz}},
  \bibinfo{author}{\bibfnamefont{W.}~\bibnamefont{Pranger}},
  \bibinfo{author}{\bibfnamefont{T.}~\bibnamefont{Antretter}},
  \bibinfo{author}{\bibfnamefont{F.~D.} \bibnamefont{Fischer}},
  \bibnamefont{and} \bibinfo{author}{\bibfnamefont{H.~P.}
  \bibnamefont{Karnthaler}}, \bibinfo{journal}{Mater. Sci. Eng. A}
  \textbf{\bibinfo{volume}{481}}, \bibinfo{pages}{479}
  (\bibinfo{year}{2008}{\natexlab{a}}).

\bibitem[{\citenamefont{5}}]{waiant:08}
\bibinfo{author}{\bibfnamefont{T.}~\bibnamefont{Waitz}},
  \bibinfo{author}{\bibfnamefont{T.}~\bibnamefont{Antretter}},
  \bibinfo{author}{\bibfnamefont{F.~D.} \bibnamefont{Fischer}},
  \bibnamefont{and} \bibinfo{author}{\bibfnamefont{H.~D.}
  \bibnamefont{Karnthaler}}, \bibinfo{journal}{Mater. Sci. Technol.}
  \textbf{\bibinfo{volume}{24}}, \bibinfo{pages}{934}
  (\bibinfo{year}{2008}{\natexlab{b}}).

\bibitem[{\citenamefont{6}}]{liucan:06}
\bibinfo{author}{\bibfnamefont{H.~B.} \bibnamefont{Liu}},
  \bibinfo{author}{\bibfnamefont{G.}~\bibnamefont{Canizal}},
  \bibinfo{author}{\bibfnamefont{P.~S.} \bibnamefont{Schabes-Retchkiman}},
  \bibnamefont{and} \bibinfo{author}{\bibfnamefont{J.~A.}
  \bibnamefont{Ascencio}}, \bibinfo{journal}{J. Phys. Chem. B}
  \textbf{\bibinfo{volume}{110}}, \bibinfo{pages}{12333}
  (\bibinfo{year}{2006}).

\bibitem[{\citenamefont{7}}]{cascue:08}
\bibinfo{author}{\bibfnamefont{A.~T.} \bibnamefont{Castro}},
  \bibinfo{author}{\bibfnamefont{E.~L.} \bibnamefont{Cuellar}},
  \bibinfo{author}{\bibfnamefont{U.~O.} \bibnamefont{Mendez}},
  \bibnamefont{and} \bibinfo{author}{\bibfnamefont{M.~J.}
  \bibnamefont{Yacaman}}, \bibinfo{journal}{Mater. Sci. Eng. A}
  \textbf{\bibinfo{volume}{481}}, \bibinfo{pages}{476} (\bibinfo{year}{2008}).

\bibitem[{\citenamefont{8}}]{fushe:04}
\bibinfo{author}{\bibfnamefont{Y.}~\bibnamefont{Fu}} \bibnamefont{and}
  \bibinfo{author}{\bibfnamefont{C.}~\bibnamefont{Shearwood}},
  \bibinfo{journal}{Scr. Mater.} \textbf{\bibinfo{volume}{50}},
  \bibinfo{pages}{319} (\bibinfo{year}{2004}).

\bibitem[{\citenamefont{9}}]{wankom:05}
\bibinfo{author}{\bibfnamefont{D.}~\bibnamefont{Wan}} \bibnamefont{and}
  \bibinfo{author}{\bibfnamefont{K.}~\bibnamefont{Komvopoulos}},
  \bibinfo{journal}{J. Mater. Res.} \textbf{\bibinfo{volume}{20}},
  \bibinfo{pages}{1606} (\bibinfo{year}{2005}).

\bibitem[{\citenamefont{10}}]{fuzha:06}
\bibinfo{author}{\bibfnamefont{Y.~Q.} \bibnamefont{Fu}},
  \bibinfo{author}{\bibfnamefont{S.}~\bibnamefont{Zhang}},
  \bibinfo{author}{\bibfnamefont{M.~J.} \bibnamefont{Wu}},
  \bibinfo{author}{\bibfnamefont{W.~M.} \bibnamefont{Huang}},
  \bibinfo{author}{\bibfnamefont{H.~J.} \bibnamefont{Du}},
  \bibinfo{author}{\bibfnamefont{J.~K.} \bibnamefont{Luo}},
  \bibinfo{author}{\bibfnamefont{A.~J.} \bibnamefont{Flewitt}},
  \bibnamefont{and} \bibinfo{author}{\bibfnamefont{W.~I.} \bibnamefont{Milne}},
  \bibinfo{journal}{Thin Solid Films} \textbf{\bibinfo{volume}{515}},
  \bibinfo{pages}{80} (\bibinfo{year}{2006}).

\bibitem[{\citenamefont{11}}]{yemis:10}
\bibinfo{author}{\bibfnamefont{J.}~\bibnamefont{Ye}},
  \bibinfo{author}{\bibfnamefont{R.~K.} \bibnamefont{Mishra}},
  \bibinfo{author}{\bibfnamefont{A.~R.} \bibnamefont{Pelton}},
  \bibnamefont{and} \bibinfo{author}{\bibfnamefont{A.~M.} \bibnamefont{Minor}},
  \bibinfo{journal}{Acta Mater.} \textbf{\bibinfo{volume}{58}},
  \bibinfo{pages}{490} (\bibinfo{year}{2010}).

\bibitem[{\citenamefont{12}}]{kolgun:08}
\bibinfo{author}{\bibfnamefont{K.}~\bibnamefont{Kolluri}},
  \bibinfo{author}{\bibfnamefont{M.~R.} \bibnamefont{Gungor}},
  \bibnamefont{and} \bibinfo{author}{\bibfnamefont{D.}~\bibnamefont{Maroudas}},
  \bibinfo{journal}{Phys. Rev. B} \textbf{\bibinfo{volume}{78}},
  \bibinfo{pages}{195408} (\bibinfo{year}{2008}).

\bibitem[{\citenamefont{13}}]{par:06}
\bibinfo{author}{\bibfnamefont{H.~S.} \bibnamefont{Park}},
  \bibinfo{journal}{Nano Lett.} \textbf{\bibinfo{volume}{6}},
  \bibinfo{pages}{958} (\bibinfo{year}{2006}).

\bibitem[{\citenamefont{14}}]{grudel:09}
\bibinfo{author}{\bibfnamefont{M.}~\bibnamefont{Gruenwald}} \bibnamefont{and}
  \bibinfo{author}{\bibfnamefont{C.}~\bibnamefont{Dellago}},
  \bibinfo{journal}{Nano Lett.} \textbf{\bibinfo{volume}{9}},
  \bibinfo{pages}{2099} (\bibinfo{year}{2009}).

\bibitem[{\citenamefont{15}}]{kadgru:03}
\bibinfo{author}{\bibfnamefont{K.}~\bibnamefont{Kadau}},
  \bibinfo{author}{\bibfnamefont{M.}~\bibnamefont{Gruner}},
  \bibinfo{author}{\bibfnamefont{P.}~\bibnamefont{Entel}}, \bibnamefont{and}
  \bibinfo{author}{\bibfnamefont{M.}~\bibnamefont{Kreth}},
  \bibinfo{journal}{Phase Transitions} \textbf{\bibinfo{volume}{76}},
  \bibinfo{pages}{355} (\bibinfo{year}{2003}).

\bibitem[{\citenamefont{16}}]{sanurb:09.2}
\bibinfo{author}{\bibfnamefont{L.}~\bibnamefont{Sandoval}} \bibnamefont{and}
  \bibinfo{author}{\bibfnamefont{H.~M.} \bibnamefont{Urbassek}},
  \bibinfo{journal}{Nano Lett.} \textbf{\bibinfo{volume}{9}},
  \bibinfo{pages}{2290} (\bibinfo{year}{2009}).

\bibitem[{\citenamefont{17}}]{finsin:84}
\bibinfo{author}{\bibfnamefont{M.~W.} \bibnamefont{Finnis}} \bibnamefont{and}
  \bibinfo{author}{\bibfnamefont{J.~E.} \bibnamefont{Sinclair}},
  \bibinfo{journal}{Philos. Mag. A} \textbf{\bibinfo{volume}{50}},
  \bibinfo{pages}{45} (\bibinfo{year}{1984}).

\bibitem[{\citenamefont{18}}]{lailiu:00}
\bibinfo{author}{\bibfnamefont{W.~S.} \bibnamefont{Lai}} \bibnamefont{and}
  \bibinfo{author}{\bibfnamefont{B.~X.} \bibnamefont{Liu}},
  \bibinfo{journal}{J. Phys. Condens. Matter} \textbf{\bibinfo{volume}{12}},
  \bibinfo{pages}{L53} (\bibinfo{year}{2000}).

\bibitem[{\citenamefont{19}}]{mutnie:10}
\bibinfo{author}{\bibfnamefont{D.}~\bibnamefont{Mutter}} \bibnamefont{and}
  \bibinfo{author}{\bibfnamefont{P.}~\bibnamefont{Nielaba}},
  \bibinfo{journal}{Phys. Rev. B} \textbf{\bibinfo{volume}{82}},
  \bibinfo{pages}{224201} (\bibinfo{year}{2010}).

\bibitem[{\citenamefont{20}}]{prokor:04}
\bibinfo{author}{\bibfnamefont{S.~D.} \bibnamefont{Prokoshkin}},
  \bibinfo{author}{\bibfnamefont{A.~V.} \bibnamefont{Korotitskiy}},
  \bibinfo{author}{\bibfnamefont{V.}~\bibnamefont{Brailovski}},
  \bibinfo{author}{\bibfnamefont{S.}~\bibnamefont{Turenne}},
  \bibinfo{author}{\bibfnamefont{I.~Y.} \bibnamefont{Khmelevskaya}},
  \bibnamefont{and} \bibinfo{author}{\bibfnamefont{I.~B.}
  \bibnamefont{Trubitsyna}}, \bibinfo{journal}{Acta Mater.}
  \textbf{\bibinfo{volume}{52}}, \bibinfo{pages}{4479} (\bibinfo{year}{2004}).

\bibitem[{\citenamefont{21}}]{hatkon:09.2}
\bibinfo{author}{\bibfnamefont{N.}~\bibnamefont{Hatcher}},
  \bibinfo{author}{\bibfnamefont{O.~Y.} \bibnamefont{Kontsevoi}},
  \bibnamefont{and} \bibinfo{author}{\bibfnamefont{A.~J.}
  \bibnamefont{Freeman}}, \bibinfo{journal}{Phys. Rev. B}
  \textbf{\bibinfo{volume}{80}}, \bibinfo{pages}{144203}
  (\bibinfo{year}{2009}).

\bibitem[{\citenamefont{22}}]{huaack:03}
\bibinfo{author}{\bibfnamefont{X.}~\bibnamefont{Huang}},
  \bibinfo{author}{\bibfnamefont{G.~J.} \bibnamefont{Ackland}},
  \bibnamefont{and} \bibinfo{author}{\bibfnamefont{K.~M.} \bibnamefont{Rabe}},
  \bibinfo{journal}{Nature Mater.} \textbf{\bibinfo{volume}{2}},
  \bibinfo{pages}{307} (\bibinfo{year}{2003}).

\bibitem[{\citenamefont{23}}]{swoand:82}
\bibinfo{author}{\bibfnamefont{W.~C.} \bibnamefont{Swope}},
  \bibinfo{author}{\bibfnamefont{H.~C.} \bibnamefont{Andersen}},
  \bibinfo{author}{\bibfnamefont{P.~H.} \bibnamefont{Berens}},
  \bibnamefont{and} \bibinfo{author}{\bibfnamefont{K.~R.}
  \bibnamefont{Wilson}}, \bibinfo{journal}{J. Chem. Phys.}
  \textbf{\bibinfo{volume}{76}}, \bibinfo{pages}{637} (\bibinfo{year}{1982}).

\bibitem[{\citenamefont{24}}]{nos:84}
\bibinfo{author}{\bibfnamefont{S.}~\bibnamefont{Nos\'{e}}}, \bibinfo{journal}{J.
  Chem. Phys.} \textbf{\bibinfo{volume}{81}}, \bibinfo{pages}{511}
  (\bibinfo{year}{1984}).

\bibitem[{\citenamefont{25}}]{kadent:02}
\bibinfo{author}{\bibfnamefont{K.}~\bibnamefont{Kadau}} \bibnamefont{and}
  \bibinfo{author}{\bibfnamefont{P.}~\bibnamefont{Entel}},
  \bibinfo{journal}{Phase Transitions} \textbf{\bibinfo{volume}{75}},
  \bibinfo{pages}{59} (\bibinfo{year}{2002}).

\bibitem[{\citenamefont{26}}]{meyent:98}
\bibinfo{author}{\bibfnamefont{R.}~\bibnamefont{Meyer}} \bibnamefont{and}
  \bibinfo{author}{\bibfnamefont{P.}~\bibnamefont{Entel}},
  \bibinfo{journal}{Phys. Rev. B} \textbf{\bibinfo{volume}{57}},
  \bibinfo{pages}{5140} (\bibinfo{year}{1998}).

\bibitem[{\citenamefont{27}}]{luoahr:03}
\bibinfo{author}{\bibfnamefont{S.-N.} \bibnamefont{Luo}},
  \bibinfo{author}{\bibfnamefont{T.~J.} \bibnamefont{Ahrens}},
  \bibinfo{author}{\bibfnamefont{T.}~\bibnamefont{Cagin}},
  \bibinfo{author}{\bibfnamefont{A.}~\bibnamefont{Strachan}},
  \bibinfo{author}{\bibfnamefont{W.~A.} \bibnamefont{Goddard~III}},
  \bibnamefont{and} \bibinfo{author}{\bibfnamefont{D.~C.} \bibnamefont{Swift}},
  \bibinfo{journal}{Phys. Rev. B} \textbf{\bibinfo{volume}{68}},
  \bibinfo{pages}{134206} (\bibinfo{year}{2003}).

\bibitem[{\citenamefont{28}}]{luoahr:03.2}
\bibinfo{author}{\bibfnamefont{S.-N.} \bibnamefont{Luo}} \bibnamefont{and}
  \bibinfo{author}{\bibfnamefont{T.~J.} \bibnamefont{Ahrens}},
  \bibinfo{journal}{Appl. Phys. Lett.} \textbf{\bibinfo{volume}{82}},
  \bibinfo{pages}{1836} (\bibinfo{year}{2003}).

\bibitem[{\citenamefont{29}}]{qinche:01}
\bibinfo{author}{\bibfnamefont{W.}~\bibnamefont{Qin}} \bibnamefont{and}
  \bibinfo{author}{\bibfnamefont{Z.~H.} \bibnamefont{Chen}},
  \bibinfo{journal}{J. Alloys Compd.} \textbf{\bibinfo{volume}{322}},
  \bibinfo{pages}{286} (\bibinfo{year}{2001}).

\bibitem[{\citenamefont{30}}]{menzho:02}
\bibinfo{author}{\bibfnamefont{Q.}~\bibnamefont{Meng}},
  \bibinfo{author}{\bibfnamefont{N.}~\bibnamefont{Zhou}},
  \bibinfo{author}{\bibfnamefont{Y.}~\bibnamefont{Rong}},
  \bibinfo{author}{\bibfnamefont{S.}~\bibnamefont{Chen}},
  \bibinfo{author}{\bibfnamefont{T.~Y.} \bibnamefont{Hsu}}, \bibnamefont{and}
  \bibinfo{author}{\bibfnamefont{Z.}~\bibnamefont{Xu}}, \bibinfo{journal}{Acta
  Mater.} \textbf{\bibinfo{volume}{50}}, \bibinfo{pages}{4563}
  (\bibinfo{year}{2002}).

\bibitem[{\citenamefont{31}}]{khaami:09}
\bibinfo{author}{\bibfnamefont{J.}~\bibnamefont{Khalil-Allafi}}
  \bibnamefont{and}
  \bibinfo{author}{\bibfnamefont{B.}~\bibnamefont{Amin-Ahmadi}},
  \bibinfo{journal}{J. Alloys Compd.} \textbf{\bibinfo{volume}{487}},
  \bibinfo{pages}{363} (\bibinfo{year}{2009}).

\bibitem[{\citenamefont{32}}]{wenhua:06}
\bibinfo{author}{\bibfnamefont{C.}~\bibnamefont{Wen}},
  \bibinfo{author}{\bibfnamefont{B.}~\bibnamefont{Huang}},
  \bibinfo{author}{\bibfnamefont{Z.}~\bibnamefont{Chen}}, \bibnamefont{and}
  \bibinfo{author}{\bibfnamefont{Y.}~\bibnamefont{Rong}},
  \bibinfo{journal}{Mat. Sci. Eng. A} \textbf{\bibinfo{volume}{438-440}},
  \bibinfo{pages}{420} (\bibinfo{year}{2006}).

\bibitem[{\citenamefont{33}}]{paw:09}
\bibinfo{author}{\bibfnamefont{P.}~\bibnamefont{Pawlow}}, \bibinfo{journal}{Z.
  Phys. Chem. Stoechiom. Verwandtschafts.} \textbf{\bibinfo{volume}{65}},
  \bibinfo{pages}{1,545} (\bibinfo{year}{1909}).


\end{thebibliography}
\end{document}